\documentclass[prb,twocolumn,showpacs,groupedaddress]{revtex4}%
\usepackage{amsmath}
\usepackage{graphicx}
\usepackage{amsfonts}
\usepackage{amssymb}%
\begin{document}
\title{The temperature influence on quantum tunneling in the spin-boson model}
\author{Zhi-De Chen}\email[Author to whom correspondence should be addressed; ]{tzhidech@jnu.edu.cn}
\author{Ying-Hua Huang}
\author{Hang Wong}
\affiliation{Department of Physics, Jinan University, Guangzhou
510632, China}

\begin{abstract}
The existing studies of the temperature effect of the spin-boson
model show that the tunneling splitting will increase with
temperature, no matter how weak the couplings are between the bath
and the spin. However, the small polaron theory said that in the
weak coupling and low temperature regime, the tunneling is, in fact,
dominated by the diagonal transitions whereas this diagonal
contribution to the tunneling should be faded away with increasing
temperature. Taking advantage of the analysis originated from the
Feynman-Vernon's influence functional theory, the influence on the
tunneling by the phonon bath can be expressed as a product of the
effects of every single phonon mode, which can be studied by
numerical exact diagonalization. We find that, in the weak coupling
and low temperature regime, all the spin-single-mode systems exhibit
the same decreasing dependence of tunneling on increasing
temperature. In turn, with the conclusion of influence functional
theory, the tunneling of the spin-boson model should decrease and it
is independent of the bath structure. In the strong coupling regime,
however, the temperature effect reverses from suppressing to
enhancing the tunneling with the increase of
temperature. Discrepancies between the old theories and the small polaron theory are also explained. 

\end{abstract}
 \maketitle

\section{Introduction}
The spin-boson model which consists of a two-state system (TSS)
coupled linearly to a phonon bath is a paradigm model of quantum
dissipative systems. The study on this model has a long history
and it still attracts a great deal of interest.$^{1-19}$ The basic
question is how thermal fluctuations from the environment
influence quantum tunneling of the TSS since all the real systems
are at finite temperature. The answer to this question is
important for the observation of macroscopic quantum phenomena and
also for the realization of the quantum computer where the
stability of a ``qubit" against thermal fluctuations is
critical.\cite{mss} Up to now, this issue has been addressed by
various treatments,\cite{weiss,bm,su,keh,ct} including numerical
calculation,\cite{rr} and it seems that all got the same
conclusion, i.e., quantum tunneling is enhanced by the temperature
in low temperature regions. However, besides the different
quantitative results, the preconditions to the conclusion are also
different. Using a perturbation approach combined with a
renormalization group (RG) treatment based on path integral, Bray
and Moore firstly found that quantum tunneling increases with
temperature provided the coupling to the bath {\it exceeds} some
critical value.\cite{bm} Numerical calculation based on
path-integral also shown that the same conclusion can be found in
the strong coupling regime but it was also stated that the
situation {\it in weak coupling regime is qualitatively different
from that in the strong coupling regime}.\cite{rr} On the other
hand, the flow equation analysis\cite{keh} and the variational
calculation\cite{sh,ct} suggested the enhancement of quantum
tunneling happens in the weak coupling and the low temperature
regions. Notably a quantitative result was derived and physics
analysis on the enhancing mechanism was also provided in the
variational calculation for the sub-Ohmic case.\cite{ct} In fact,
the quantitative result found in the sub-Ohmic case is in
agreement with that found by Weiss,\cite{weiss} such a $T^{1+s}$
dependence was also considered as the universality in dissipative
two-state system and was declared to be valid for all the coupling
strength before the
localization happens.\cite{su} 
Nevertheless, physics analysis from the well-known small polaron
theory tells a different story.

It is known that the zero-biased spin-boson model is just a simple
example of the polaron-phonon (or exciton-phonon) system, i.e., the
two-site problem.\cite{ss} According to the small polaron
theory,\cite{mah} the contribution to tunneling has two parts at
finite temperature, say, the diagonal and non-diagonal contribution.
In the weak coupling regime, the diagonal part makes the main
contribution to quantum tunneling and hence quantum tunneling should
decrease with increasing temperature in the low temperature regions,
a result that is in confliction with some
analysis of the spin-boson model mentioned above. 
The discrepancy between various treatments indicates that the
consensus of temperature effect is still lacking and this is the
main interest of the present work.

In this paper, the temperature effect on quantum tunneling in the
spin-boson model is studied by analysis originated from the
influence functional theory developed by Feynman and
Vernon.\cite{fv} One important conclusion of this theory is that
the contribution of the whole bath can be expressed as a weighted
integration over the contribution of single phonon modes; hence
the temperature effect on quantum tunneling can be qualitatively
known by studying the temperature effect in the single mode case
which can be exactly done by numerical calculation. It is found
that, in the weak coupling regime, all the phonon modes contribute
the same temperature effect, i.e., quantum tunneling is suppressed
by temperature, and the enhancing mechanisms by both the
variational calculation\cite{ct} and flow equation
analysis\cite{keh} do not exist in a real system. In the strong
coupling regime, however, the temperature effect is reversed, in
agreement with the known results.\cite{bm,rr} The organization for
the rest of the paper is as follows. In the next section, the
model and the explanation of our calculation method are presented.
Numerical results for both $T=0$ and $T\not=0$ cases are given in
Sec.\ III. Conclusion and discussion are presented in the last
section.

\section{The model and the explanation}
The Hamiltonian of the spin-boson model is given by (setting
$\hbar=1$)\cite{leg,weiss}
\begin{equation}
\hat{H}=-\frac{\Delta_0}{2}\hat{\sigma}_x+\sum_k\omega_k
\hat{b}_k^{\dagger}\hat{b}_k+\hat{\sigma}_z\sum_k
g_k(\hat{b}_k^{\dagger}+\hat{b}_k),
\end{equation}
where $\hat{\sigma}_i(i=x,y,z)$ is the Pauli matrix,
$\hat{b}_k(\hat{b}_k^{\dagger})$ is the annihilation (creation)
operator of the $k$th phonon mode with energy $\omega_k$ and $g_k$
is the coupling parameter. The solution of this model depends on the
so-called bath spectral function (density) which is defined
as\cite{leg,weiss} $ J(\omega)=\pi
\sum_kg_k^2\delta(\omega-\omega_k). $ For frequencies below a
high-energy cutoff $\omega_c$, the spectral function has the power
law form,
\begin{equation}
J(\omega)=\frac{\pi}{2}\alpha
\omega^s/\omega_c^{s-1},~~~~~~~0<\omega\le\omega_c,
\end{equation}
where $\alpha$ is a dimensionless coupling strength which
characterizes the dissipation strength and the parameter $s$
characterizes the property of the bath, i.e., $0<s<1$, $s=1$, and
$s>1$ represent, respectively, the sub-Ohmic, Ohmic, and super-Ohmic
cases. It should be noted that the solution of the model (1) is
merely determined by the spectral function $ J(\omega)$ comes from
the fact that the bath degree of freedom can be integrated out as
Gaussian integrals,\cite{se,ck} an ideas originated from the
influence functional theory.\cite{fv}
\begin{figure}[h]
\centering
\includegraphics[width=0.5\textwidth]{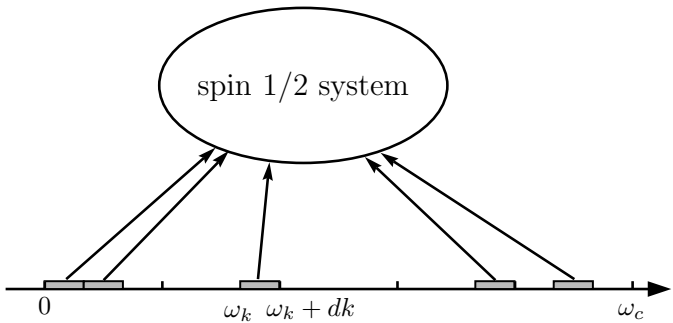}
\caption{Illustration of how the phonon bath is divided into a
series of dynamically and statistically independent subsystems in
continuum limit. Each subsystem represents a single-mode system for
modes with frequencies lie in $(\omega_k,\omega_k+d\omega_k]$.}
\end{figure}

As shown in Ref.\ \onlinecite{fv}, the phonon bath can be
considered as a distribution of single modes and the single modes
can be treated as dynamically and statistically independent
subsystems. An illustration of how the phonon bath with frequency
lies in $(0,\omega_c]$ is divided in to a series of sub-systems,
i.e., the single-mode systems, is shown in Fig.\ 1. According to
the influence functional theory,\cite{fv} the total influence
functional of the bath can be expressed as a product of that of
each subsystem
\begin{equation}
{\cal F}=\prod_k{\cal F}_k,
\end{equation}
where ${\cal F}_k$ is the influence functional of the $k$th mode.
The validity of the above result is based on the assumption that the
correlation between different modes can be ignored even when the
interaction between the TSS and phonon modes is considered. This
implies that the ground state of the whole system can be expressed
as
\begin{equation}
|\Phi_0\rangle=\frac{1}{\sqrt{2}}\sum_{\pm}|\pm\rangle
\prod_k|\phi_k^{\pm}(0)\rangle,~~~~~~~~
\langle\phi_k^{\pm}(0)|\phi_{k'}^{\pm}(0)\rangle=\delta_{kk'},
\end{equation}
where $|\pm\rangle$ is the eigen-state of $\hat{\sigma}_z$ and
$|\phi_k^{\pm}(0)\rangle$ is the eigen-state of the $k$th mode when
it interacts with the TSS {\it individually}. Accordingly, one can
find out the tunneling splitting
\begin{equation}
\Delta/\Delta_0=\langle\Phi_0|\hat{\sigma}_x|\Phi_0\rangle=\prod_k
d_k(0),
\end{equation}
where $d_k(0)=\langle\phi^+_k(0)|\phi^-_k(0)\rangle$ is the
dressing factor contributed by the $k$th mode.

The above result can be generalized to the $T\not=0$ case by
assuming that all the subsystems are still statistically independent
when the interactions with the TSS are considered. In the influence
functional theory, a trace over each phonon mode is made in Eq.\
(3).\cite{fv} In the present case, the generalization can be done in
two steps. Firstly, all the excited states of the whole system are
suggested to have the same form as the ground state given in Eq.\
(4), that is, the $n$th excited state with eigenvalue $E_n$ can be
expressed as
\begin{equation}
|\Phi_n\rangle=\frac{1}{\sqrt{2}}\sum_{\pm}|\pm\rangle
\prod_k|\phi_k^{\pm}(n)\rangle,
\end{equation}
where
$\langle\phi_k^{\pm}(n)|\phi_{k'}^{\pm}(n')\rangle=\delta_{kk'}\delta_{nn'}$,
then the tunneling splitting at $T\not=0$ can be found by
\begin{eqnarray}
\Delta_T/\Delta_0&=\frac{1}{Z}\sum_ne^{-\beta
E_n}\langle\Phi_n|\hat{\sigma}_x|\Phi_n\rangle\\
\nonumber
 &=\frac{1}{Z}\sum_ne^{-\beta E_n}\prod_k~d_k(n),
\end{eqnarray}
where
\begin{equation}
Z=\sum_ne^{-\beta
E_n},~~~~~~~~~d_k(n)=\langle\phi_k^{\mp}(n)|\phi_k^{\pm}(n)\rangle.
\end{equation}
The assumption that each subsystem is still statistically
independent means
\begin{equation}
Z=\prod_k z_k,
\end{equation}
where $z_k=\sum_ne^{-\beta e_n(k)}$ is the partition function of the
$k$th mode interacting with the TSS individually. Therefore, one can
find
\begin{equation}
\Delta_T/\Delta_0=\prod_k~D_k(T),~~~~~D_k(T)=\frac{1}{z_k}\sum_ne^{-\beta
e_n(k)}d_k(n).
\end{equation}
The above analysis shows that, by Eqs.\ (5) and (10), the
contribution of the phonon bath to the tunneling splitting is a
product of the contribution from every single mode. This is the
starting point of our calculation. In the next section, the
validity of such a simple product assumption (i.e., Eqs.\ (5) and
(10)) will be verified numerically in a simple way.

Although other analytical treatments, such as the variational
calculation and flow equation method, are not based on the influence
functional theory, the expression of tunneling splitting found have
exactly the same forms given in Eqs.\ (5) and
(10).  
In fact, the dressing factor of a single mode can be expressed as
the form of influence functional given in Ref.\ \onlinecite{fv},
\begin{equation}
D_k(T)=\exp\{-\psi_k(T)\},
\end{equation}
where $\psi_k(T)$ can be considered as the dressing phase of the
$k$th mode. Since the solution should be determined by the
spectral function only, a natural choice of 
the dressing phase is
\begin{equation}
\psi_k(T)=g_k^2f(\omega_k, T),
\end{equation}
which is just the form found by other analytical treatments,
including the perturbation calculation, the variational calculation,
and the flow equation method in spite of their self-consistent
forms.\cite{sh,ct,keh,zz,st} This implies that the ideas of
influence functional theory has been tacitly adopted by analytical
treatments generally. It should be noted that, in the present work,
the dressing factor $D_k(T)$ is calculated numerically, and hence
one can only get qualitative conclusion and no quantitative result
can be found.
\begin{figure*}
\centering
\begin{minipage}[c]{0.8\textwidth}
\centering
\includegraphics[width=\textwidth,height=4.1in]{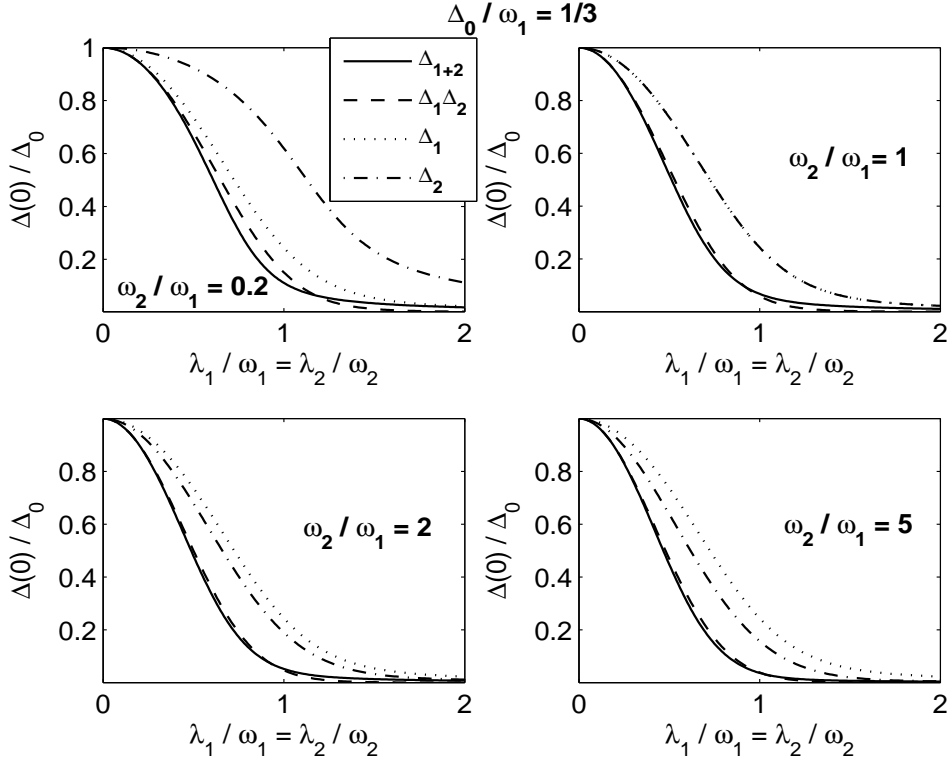}
\end{minipage}
\begin{minipage}[c]{0.19\textwidth}
\centering \caption{Numerical verification of the simple product
assumption in the case of $T=0$ and $\Delta_0/\omega_{1}= 1/3$.
The deviation between $\Delta_{1+2}$ and $\Delta_1\Delta_2$
appears as the coupling strength increases but both $\Delta_{1+2}$
and $\Delta_1\Delta_2$ show similar
$\lambda/\omega_{1,2}$-dependence. In the weak coupling regime
with $\lambda/\omega_{1,2}\le 0.1$, it is found that
$\Delta_{1+2}=\Delta_1\Delta_2$ holds well within calculation
error.}
\end{minipage}%
\end{figure*}
\begin{figure*}
\centering
\begin{minipage}[c]{0.8\textwidth}
\centering
\includegraphics[width=\textwidth,height=4.1in]{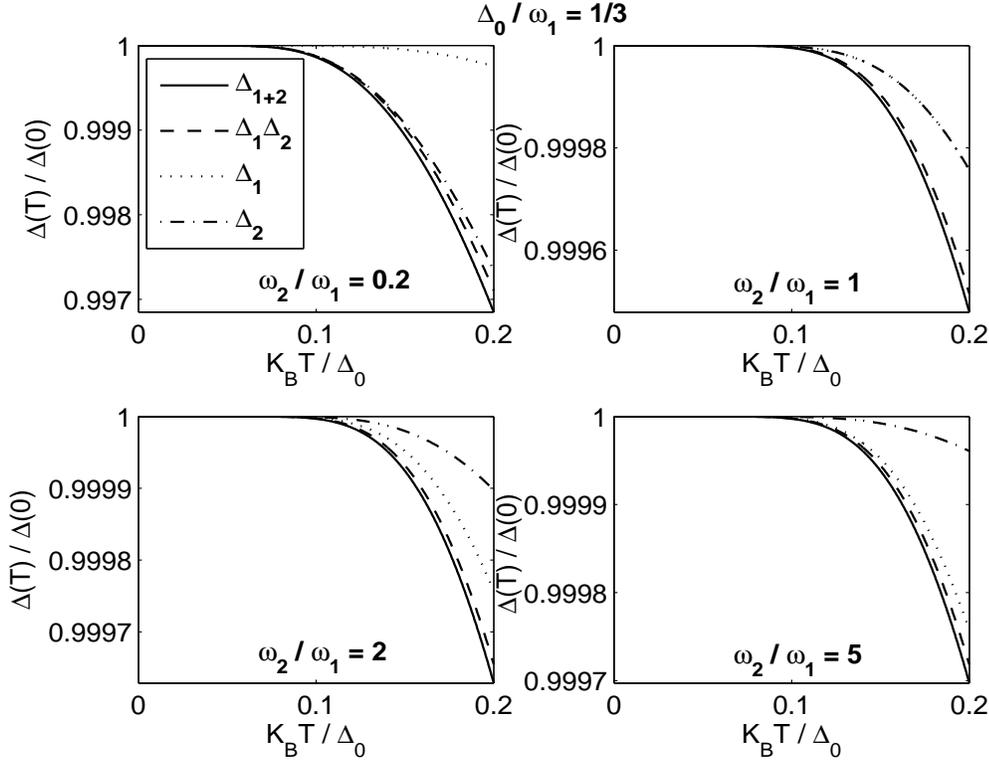}
\end{minipage}
\begin{minipage}[c]{0.19\textwidth}
\centering \caption{Numerical verification of the simple product
assumption in the case of $T\neq0$ with $\lambda/\omega_{1,2}=
0.1$ and $\Delta_0/\omega_{1}= 1/3$. The deviation between
$\Delta_{1+2}(T)$ and $\Delta_1(T)\Delta_2(T)$ becomes obvious as
temperature increases but both $\Delta_{1+2}(T)$ and
$\Delta_1(T)\Delta_2(T)$ always show similar temperature
dependence. In the low temperature regions of $k_BT/\Delta_0\le
0.1$, it is found that $\Delta_{1+2}(T)=\Delta_1(T)\Delta_2(T)$
holds well in the weak coupling regime with $\lambda/\omega_{1,2}<
0.1$.}
\end{minipage}%
\end{figure*}

\section{Temperature effect on quantum tunneling}
According to the above analysis, the influence of the bath on
quantum tunneling can be known by solving the following single-mode
system
\begin{equation}
\hat{h}=-\frac{\Delta_0}{2}\hat{\sigma}_x+\omega
\hat{b}^{\dagger}\hat{b}+\hat{\sigma}_z\lambda
(\hat{b}^{\dagger}+\hat{b}),
\end{equation}
where $\lambda/\omega$ is a dimensionless coupling parameter. The
above Hamiltonian is solved by numerical diagonalization using the
basis $|\pm\rangle\otimes \{|n\rangle\}$, here $|n\rangle$ is the
eigenstate of $\hat{b}^{\dagger}\hat{b}$ and $n=0,1,2,\cdots$. To
find out the dressing factor $D_k(T)$ in Eq.\ (10), we shall take
the first $N$ eigenvalues $\{e_n\}$ and the corresponding
eigenstates $\{\varphi_n\}$ for calculation in practice, here $N$ is
determined by
\begin{equation}
\beta(e_N-e_1)\ge L,
\end{equation}
and $L$ is a fixed number to control the calculation error. It is
found that the result becomes $N$-independent when $L\ge 20$.
Since the sign of
$\langle\varphi_n|\hat{\sigma}_x|\varphi_n\rangle$ is not relevant
for the calculation of the total tunneling splitting, we have
\begin{equation}
D_k(T)\simeq \frac{1}{z}\sum_{n=1}^N~e^{-\beta
e_n}|\langle\varphi_n|\hat{\sigma}_x|\varphi_n\rangle|,
\end{equation}
where $z\simeq \sum_{n=1}^Ne^{-\beta e_n}$. We have compared
numerical result with other approach using a different
basis,\cite{ir} and the result is the same within the calculation
error. A detail discussion on numerical diagonalization of the
single mode Hamiltonian can be found in Ref.\ \onlinecite{huang}.

In the following, we shall present a simple verification of Eqs.\
(5) and (10) by numerical calculation. This is done in the
following way. Firstly, the numerical solutions for two given
modes with frequencies $\omega_1$ and $\omega_2$ and coupling
parameters $\lambda_1/\omega_1$ and $\lambda_2/\omega_2$ are found
separatively, and the tunneling splitting of the ground state
$\Delta_{1,2}$ and the temperature dependence of $\Delta_{1,2}(T)$
can be found by Eq.\ (15). Secondly, the numerical solution for a
two-mode system with the same frequencies and coupling parameters
are done in the same way, and the tunneling splitting of the
ground state $\Delta_{1+2}$ and the temperature dependence
$\Delta_{1+2}(T)$ can be found. According to Eqs.\ (5) and (10),
one should have
\begin{equation}
\Delta_{1+2}=\Delta_1\Delta_2,~~~~{\rm
and}~~~\Delta_{1+2}(T)=\Delta_1(T)\Delta_2(T).
\end{equation}
Typical result for the ground state (i.e., the case of $T=0$) is
shown in Fig.\ 2. It is seen that the validity of the above equation
depends on parameters $\omega_1/\omega_2$, $\omega_{1,2}/\Delta_0$,
and $\lambda/\omega_{1,2}$. However, one can find that the above
equation holds quite well within the calculation errors when
$\lambda/\omega_{1,2}\le 0.1$. The result for the case of $T\not=0$
is shown in Fig.\ 3. One can see that, in this case, the validity of
the above equation also depends on the temperature region. The
$\Delta_{1+2}(T)=\Delta_1(T)\Delta_2(T)$ holds for
$\lambda/\omega_{1,2}< 0.1$ in the temperature regions with
$k_BT/\Delta_0<0.1$. Since in a real thermodynamical system, the
coupling parameter of each single mode should scale with $1/N$ and
$N\rightarrow \infty$, one can conclude that Eqs.\ (5) and (10) hold
in the low temperature region with $k_BT/\Delta_0<0.1$.

\begin{figure}[h]
\centering
\includegraphics[width=0.5\textwidth]{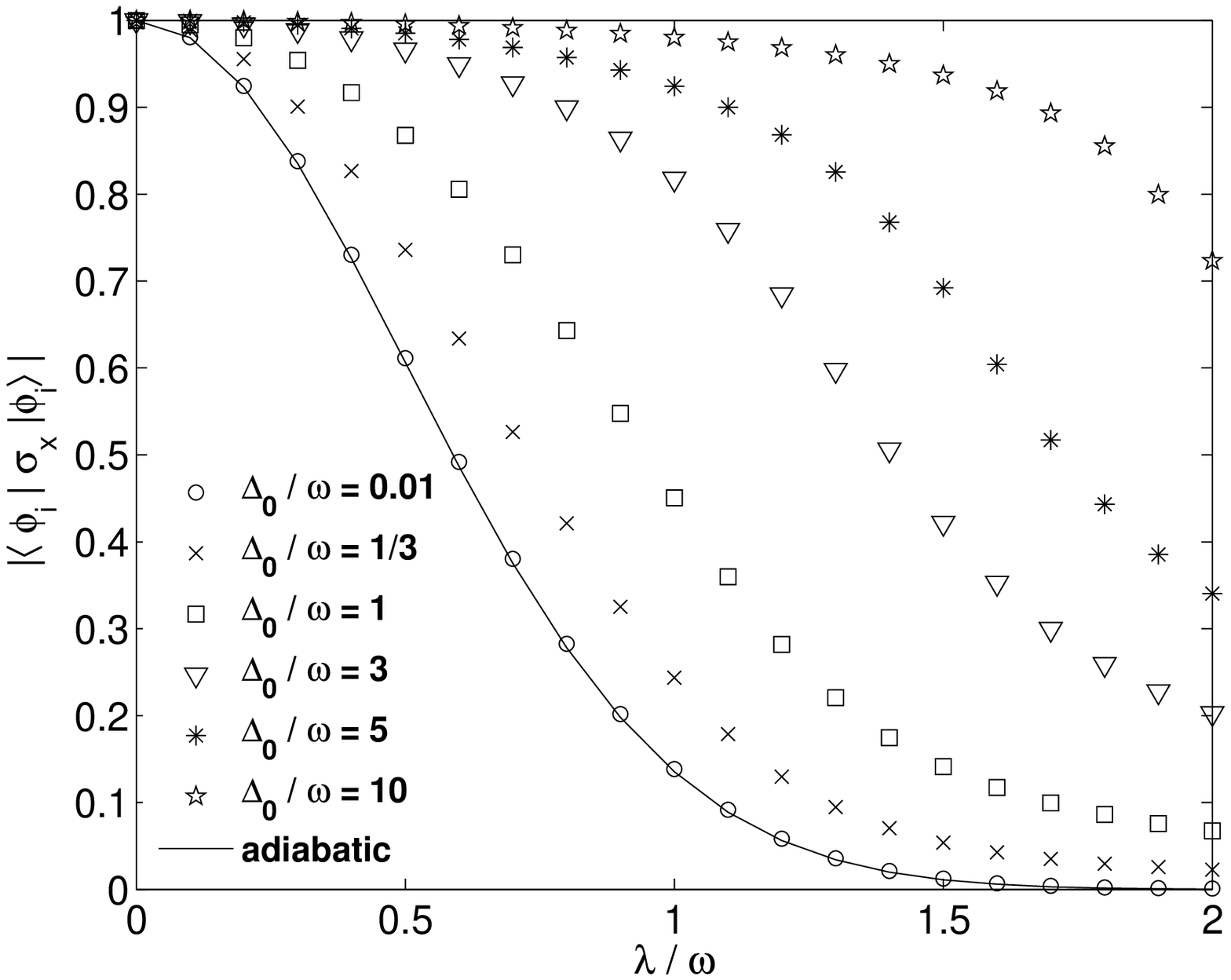}
\caption{Dependence of the tunneling splitting of the ground state
on the coupling parameter $(\lambda/\omega)$ for various
frequencies. For comparison, the result by adiabatic approximation
is shown in solid line. The numerical result agrees with the result
by adiabatic approximation only when $\Delta_0/\omega\le 0.1$, and
the lower the frequency the larger the discrepancy. }
\end{figure}

The above analysis shows that numerical solutions to the single-mode
systems can serve as the solid basis for understanding the effect of
the whole bath. As we have shown, tunneling splitting found by
analytical treatments with various approximations have the form
given in Eq.\ (11), this implies that numerical solution of the
single-mode system can be used as a touchstone to test qualitatively
the analytical results. Let us first see the result of the ground
state. Figure 4 shows the dependence of the tunneling splitting of
the ground state on the coupling parameter for various frequencies.
For comparison, we also present the result by adiabatic
approximation, i.e., $\Delta_a/\Delta_0=\langle
0_{\pm}|0_{\mp}\rangle$, where
$|0_{\pm}\rangle=\exp\{\mp(\lambda/\omega)(\hat{b}^{\dagger}-\hat{b})\}|0\rangle$
is the displaced-oscillator ground state. It shows that the
numerical result agrees with the adiabatic approximation for
$\Delta_0/\omega\le 0.1$, but discrepancy becomes obvious when
$\Delta_0/\omega\ge 1/3$ and the lower the frequency, the larger the
deviation. The result shown in Fig.\ 4 provides a quantitative
explanation for the fact that the usual Debye-Waller reduction by
adiabatic approximation on tunneling {\it often appears to be by too
large}.\cite{zw} As one can see from Fig.\ 4, in the weak coupling
regime with $\lambda/\omega\le 0.1$, the low frequency modes with
frequency $\Delta_0/\omega\ge 5$ have almost no effect on the
tunneling splitting, while the Debye-Waller factor tells that the
low-frequency modes lead to localization.\cite{zw,ct,wh} As a matter
of fact, one goal for analytical treatment is to find an appropriate
dressing factor better than the  usual Debye-Waller
factor. Considering the dressing phase given in Eq.\ (11), 
the result by the flow equation analysis at $T=0$ is\cite{keh}
\begin{equation}
\psi_k(T=0)=-2\frac{g_k^2}{\omega_k^2-\Delta^2},
\end{equation}
where $\Delta$ is the tunneling splitting of the whole system and
was denoted as $\Delta_{\infty}$ in Ref.\ \onlinecite{keh}.
Interestingly, this result is in agreement with the perturbation
calculation up to the first order approximation with $\Delta$ is
replaced with $\Delta_0$.\cite{st} According to the above result,
the bath is demarcated by a frequency $\omega^*=\Delta$(or
$\omega^*=\Delta_0$ in perturbation calculation), the modes with
frequency larger than $\omega^*$ will suppress tunneling splitting,
while the modes with frequency lower than $\omega^*$ will enhance
tunneling splitting since the dressing phase changes its sign.
However, our numerical calculation find no evidence for the
existence of $\omega^*$. We find that all the phonon modes lowers
tunneling splitting in the coupling regime $0<(\lambda/\omega)\le
2.0$ at $T=0$. On the other hand, the variational calculation
gives\cite{zw,ct,wh2}
\begin{equation}
\psi_k(T=0)=-2\frac{g_k^2}{(\omega_k+\Delta)^2}.
\end{equation}
Comparing with the usual Debye-Waller factor, i.e.,
$\psi_k(T=0)=-2g_k^2/\omega_k^2$, the above result effectively
shifts the frequency origin from $\omega=0$ to $\omega=\Delta$, viz,
the contribution of the low frequency modes with frequency
$\omega<\Delta$ is ignored, an ideas similar to adiabatic
renormalization.\cite{leg,zw} As one can see in Fig.\ 4, such a
treatment is reasonable since the low frequency modes have less
effect on tunneling splitting in the weak coupling regime. This
explains the fact that the variational ground state is a good
approximation to the true ground state in the weak coupling
regimes.\cite{wh1} However, the variational calculation fails to
trace the phonon-induced localization in the strong coupling
regime.\cite{wh2}

\begin{figure}[h]
\centering
\includegraphics[width=0.5\textwidth]{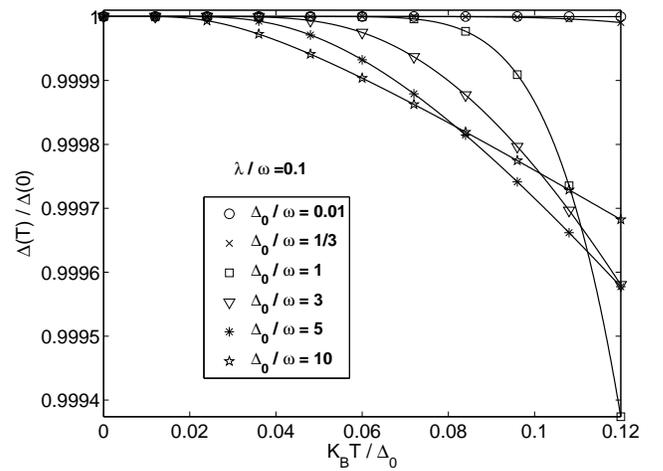}
\caption{Temperature dependence on the dressing factor for the
single-mode system in the case of $(\lambda/\omega)=0.1$ for various
frequencies. We have $d\Delta_k(T)/dT<0$ for all the phonon modes.
In the low temperature region with $k_BT/\Delta_0\le 0.1$, the
phonon modes with frequency $\Delta_0/\omega <0.1$ show almost no
observable temperature effect. }
\end{figure}

Now we turn to see the temperature effect on tunneling splitting.
Temperature dependence on the dressing factor for the single-mode
system is calculated by Eq.\ (15) and some typical results are shown
in Fig.\ 5. In the weak coupling regime, we have
$d\Delta_k(T)/dT<0$ for all the phonon modes. 
For a given coupling strength with $(\lambda/\omega)\le 0.1$, the
lower the frequency (i.e., the larger $\Delta_0/\omega$), the lower
temperature where tunneling splitting starts to decrease. This is
easy to understand since the contribution of the excited states
appears at a higher temperature for a larger frequency. 
 In the low temperature region with $k_BT/\Delta_0\le
0.1$, the phonon modes with frequency $\Delta_0/\omega <0.1$ show
almost no observable temperature effect. By Eq.\ (10), this result
shows that, in the weak coupling regimes, tunneling splitting
decreases with increasing temperature in the low temperature region
with $k_BT/\Delta_0<0.1$. It should be noted that, since all the
modes have the same temperature dependence, the result of whole bath
according to Eq.\ (10) is qualitatively independent of the weight of
different modes, namely, it holds for the sub-Ohmic, Ohmic, and
super-Ohmic cases. Obviously, this conclusion is in confliction with
some known results\cite{weiss,sh,keh,su,ct} and explanations will be
given below.

\begin{figure*}
\centering
\begin{minipage}[c]{0.8\textwidth}
\centering
\includegraphics[width=\textwidth,height=4.1in]{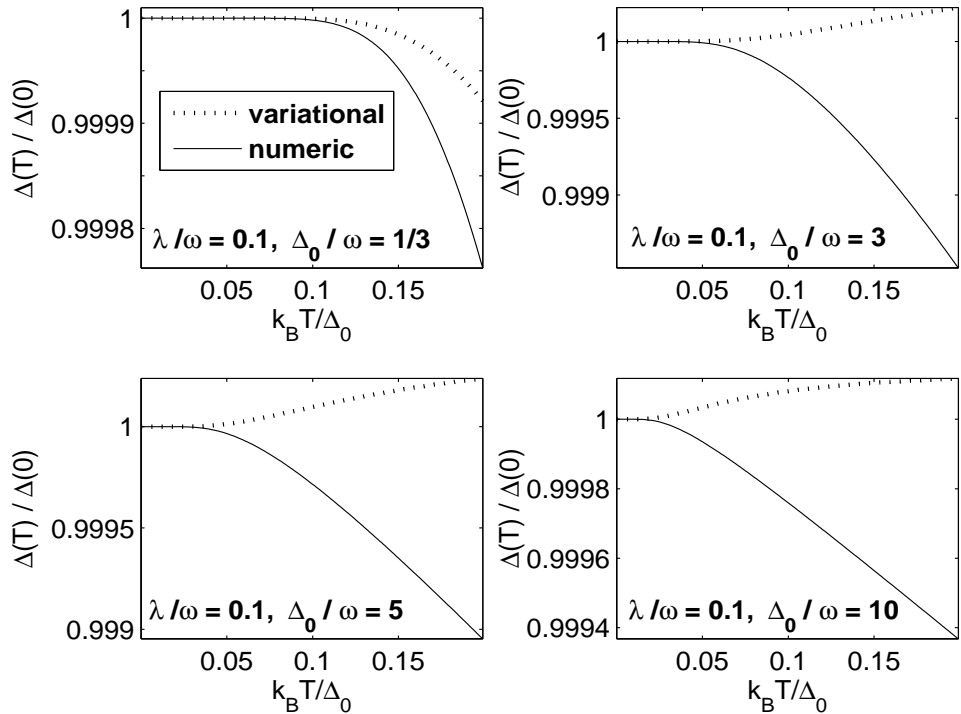}
\end{minipage}
\begin{minipage}[c]{0.19\textwidth}
\centering \caption{Temperature dependence of tunneling splitting
in single-mode case by the variational calculation (shown as
dot-line) and numerical calculation (shown as solid line). For
high frequency with $\Delta_0/\omega \le 1$, temperature
dependence is qualitatively the same for both cases that tunneling
splitting decreases with increasing temperature. However, in the
low frequency region with $\Delta_0/\omega > 1$, temperature
dependence becomes opposite for both cases, showing that the
enhancing mode is an artifact of the variational calculation.}
\end{minipage}%
\end{figure*}
According to the variational calculation,\cite{sh,ct} temperature
enhancing mechanism comes from the result that the dressing factors
for frequency modes with frequency $k_BT<\omega<\Delta$ decrease
with increasing temperature. We have searched for these enhancing
modes in numerical calculation, however,  no evidence for the
existence of such enhancing modes can be found in the weak coupling
regime. The searching work is done in the following way. We perform
a variational calculation for a single-mode system given in Eq.\
(13) by the same way as done in Ref.\ \onlinecite{ct}, then a
self-consistent equation for the tunneling splitting of the
single-mode system can be found
\begin{equation}
\Delta_v(T)/\Delta_0=\exp\{-(\lambda/\omega)^2f_1(T,\omega)\},
\end{equation}where
\begin{equation}
f_1(T,\omega)=\frac{\coth(\beta\omega/2)}{\{\omega+\Delta_v(T)\tanh[\beta\Delta_v(T)/2]
\coth(\beta\omega/2)\}^2},
\end{equation}
from which the temperature dependence on tunneling splitting can be
obtained numerically. It can be verified that, in the weak coupling
regime $(\lambda/\omega)\le 0.1$, the variational ground state is a
good approximation to the true ground state and the
temperature-enhancing modes do exist as predicted in multi-mode
case,\cite{ct} that is, we have $d\Delta_v(T)/dT>0$ when
$k_BT<\omega<\Delta_v$. Some typical results are shown in Fig.\ 6,
where one can see that, for the high frequency modes with
$\Delta_0/\omega \le 1$, the temperature dependence is qualitatively
the same for both the variational calculation and numerical
calculation that tunneling splitting decreases with increasing
temperature. However, in the low frequency region with
$\Delta_0/\omega
> 1$, the result by variational calculation shows opposite
temperature dependence to numerical result. This result indicates
that the enhancing modes could be an artifact of the variational
calculation in the case of $T\not=0$. It should also be noted that
another related prediction, the re-entrance phenomena,\cite{ct} has
been shown to be impossible in a real system.\cite{wh}

In the flow equation analysis, the enhancement of tunneling
splitting by temperature comes from the existence of a frequency
boundary $\omega^*=\Delta$. As one can see from Eq.\ (17), those
modes with frequency $\omega<\omega^*=\Delta$ always enhance the
tunneling splitting, while in the low temperature region with
$k_BT<\Delta_0$, the main contribution of the whole bath comes from
the modes with frequency $0<\omega<\Delta$, this leads to the
enhancement of tunneling splitting by temperature as shown in Ref.\
\onlinecite{keh}. However, as we have shown, there is no evidence
for the existence of $\omega^*$, hence the enhancing mechanism in
this case does not exist. We believe that, in the case of $T=0$, the
enhancement of the low frequency modes by Eq.\ (17) can compensate
for the over-estimation of the usual Debye-Waller reduction of the
high frequency modes, which leads to a good final result. In the
case of $T\not=0$, however, the fake enhancing mechanism can lead to
a wrong temperature dependence in the low temperature regions.

In the above analysis, temperature enhancing mechanisms in the weak
coupling regime by both the variational calculation and flow
equation analysis have been ruled out. Now we turn to see what
happens in the strong coupling regime. In this case, it is worth
noting that the analysis in the strong coupling regime may not be as
precisely as that in the weak coupling regime. This is because, in
the strong coupling regime, the assumption of simple product (e.g.,
Eqs.\ (5) and (10)) does not hold exactly. However, our numerical
result shows that, in the whole coupling range we studied,
$\Delta_{1+2}(T)$ always shows similar temperature dependence as
$\Delta_1(T)\Delta_2(T)$, and furthermore, to the first order
approximation (i.e., by omitting the correlation between different
modes), the effect of the whole bath can be considered as a
weighted-integration over different modes, hence the temperature
effect in the strong coupling regime can be approximately traced in
the same way as that in the weak coupling regime. In the following,
it will be shown that the temperature enhancing mechanism appears in
the strong coupling regime.

According to Eq.\ (15), if the tunneling splittings of the first
several excited states are larger than that of the ground state,
the resulted $\Delta_k(T)$ will increase with temperature,
otherwise $\Delta_k(T)$  will decreases with increasing
temperature. In other words, one can see the temperature
dependence by examining the spectrum of tunneling splitting. In
the weak coupling regime and for all the frequency ranges we
studied, $|\langle\varphi_n|\hat{\sigma}_x|\varphi_n\rangle|$
decreases with increasing $n$ for $n<5$, leading to a decreasing
tunneling splitting with temperature; however, as the coupling
parameter $(\lambda/\omega)$ increases to
some critical value, which is frequency-dependent, 
$|\langle\varphi_n|\hat{\sigma}_x|\varphi_n\rangle|$ turns to
increase with $n$ and, of course, the temperature dependence
reverses. Typical results are shown in Fig.\ 7, where one can see
the variation of the spectrum of tunneling splitting and the
reversed temperature dependence as the coupling strength exceeds
some critical value. Since the reversed temperature dependence in
the strong coupling regime is found for all the frequency range we
studied, one can conclude that tunneling splitting increases with
temperature in the spin-boson model provided that the coupling to
the bath is strong enough, confirming the results having been found
before.\cite{bm,rr}. Such an evolution of temperature dependence
with coupling to the bath is in agreement with physics analysis of
the small polaron theory. According to the small polaron
theory,\cite{mah} the contribution to the tunneling has two parts:
the diagonal and non-diagonal contribution. The diagonal
contribution decreases with increasing temperature and can only
survive in the weak coupling regime. On the other hand, the
non-diagonal contribution increases with the temperature and only
becomes important when the coupling strength is not too low. In the
weak coupling regime and low temperature region, the main
contribution to the tunneling is the diagonal part, which decreases
with increasing temperature. In the strong coupling regime, the
diagonal contribution is suppressed and non-diagonal part makes the
main contribution to the tunneling, hence tunneling splitting will
increase with temperature. In the medium coupling range, a
coherent-incoherent transition can happen when temperature increases
as shown in Ref.\ \onlinecite{huang}, where one can also find a
detail analytical analysis on temperature dependence of tunneling
splitting for single-mode system by using the small polaron theory.
The result shown in Fig.\ 7 provides an alternative explanation for
the results found in Refs.\ \onlinecite{bm} and \onlinecite{rr}.
\begin{figure*}
\centering
\begin{minipage}[c]{0.7\textwidth}
\centering
\includegraphics[width=\textwidth,height=6in]{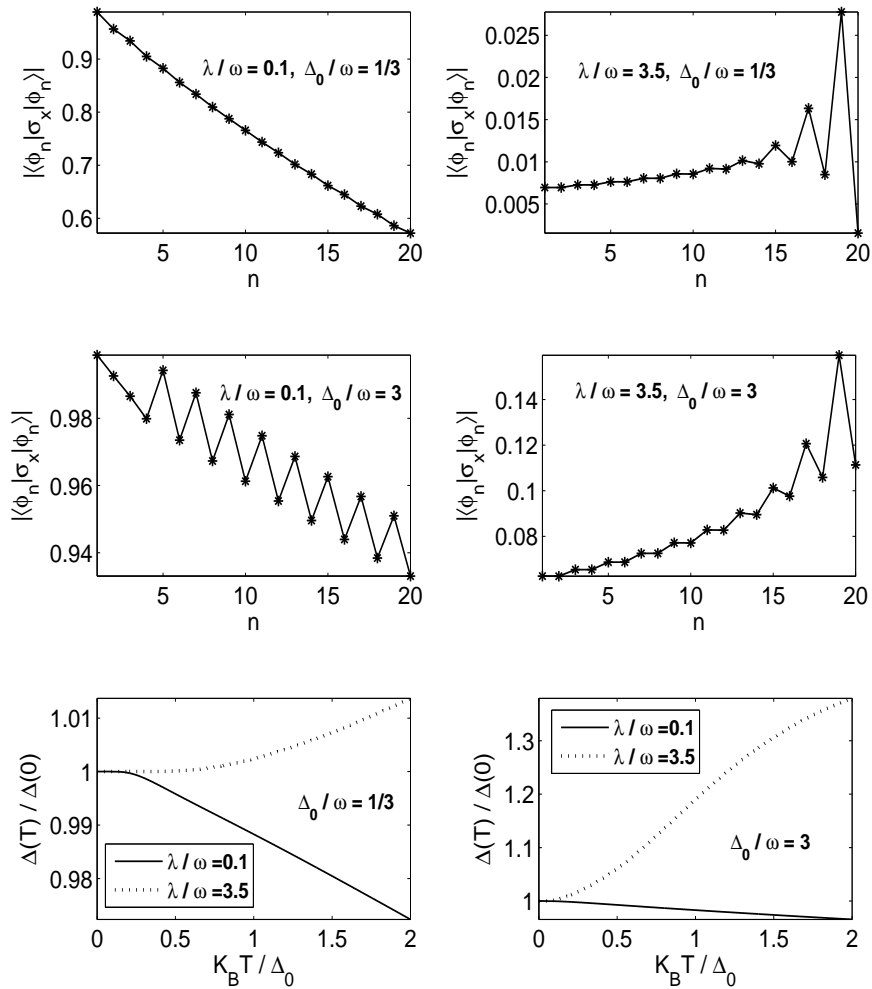}
\end{minipage}
\begin{minipage}[c]{0.25\textwidth}
\centering \caption{Spectrum of tunneling splitting for two
typical frequencies of $\Delta_0/\omega=1/3$ (up) and
$\Delta_0/\omega=3$ (medium) in the weak (left) and strong (right)
coupling regime(the line is guidence for the eyes). In weak
coupling regime,
$|\langle\varphi_n|\hat{\sigma}_x|\varphi_n\rangle|$ decreases
with $n$, while in the strong coupling regime, the dependence
reverses. The resulted temperature dependence on tunneling
splitting is shown in the lowest part, where one can see the
temperature dependence reverses in the strong coupling regime. }
\end{minipage}%
\end{figure*}

\section{Conclusion and discussion}
In the previous sections, the temperature effect on tunneling
splitting of the TSS is studied by a theory originated from the
influence functional theory. It is shown that the effect of the
whole bath can be expressed as a simple product of every single
mode, then the single-mode system is solved by numerical
diagonalization, from which the effect of the whole bath can be
found. In the weak coupling regime, all the phonon modes have the
same effect on tunneling splitting of the TSS, i.e., coupling to any
phonon modes suppresses the tunneling splitting, and the suppressing
effect increases with temperature. The result indicates that
tunneling splitting will decrease with increasing temperature for
the sub-Ohmic, Ohmic, and super-Ohmic cases. Temperature enhancing
mechanisms by both variational calculation and the flow equation
analysis have been ruled out by numerical analysis. The reversed
temperature effect on tunneling splitting in the strong coupling
regime is also explained. Our analysis shows that the temperature
effect in the spin-boson model can be understood based on the small
polaron theory. The present work can help to resolve the
discrepancy between various treatments. 
 However, in the present stage, it is not totally clear about the
conclusion found in Refs.\ \onlinecite{weiss} and \onlinecite{su}.
Here we just want to emphasize two points. First of all, the
temperature effect on quantum tunneling in the spin-boson model is
hard to deal with since the usual Debye-Waller factor tends to
zero even at $T=0$ for both the Ohmic and sub-Ohmic case. This
implies that, the diagonal contribution to tunneling dies out by
the small polaron theory and hence calculations done along this
line is impossible to include the diagonal contribution. As we
known, temperature dependence on the diagonal contribution is
opposite to that of the non-diagonal contribution. Accordingly, it
is possible to get a correct temperature dependence only when the
calculation is made by a better approximation than the small
polaron theory. Unfortunately most analytical calculations are
done in similar way to the small polaron theory. By comparing the
main result (see Eq.\ (9) in Ref.\ \onlinecite{su} and Eq.\ (3.36)
in Ref.\ \onlinecite{leg}), it is seen that the result found in
Ref.\ \onlinecite{su} is similar to that found along the line of
small polaron theory.\cite{leg} This seems to suggest that the
diagonal contribution might be omitted in the calculation of Ref.\
\onlinecite{su}. The second point is the difference between the
TSS system and the double-well system. In a double-well system,
there are excited states inside each well and the tunneling
splitting increases exponentially with the excited
state.\cite{ia,em} The temperature dependence on tunneling
splitting can be known without knowing the bath effect on
tunneling splitting of each level. As the temperature increases,
the occupation probability of the excited states increases, which
makes the tunneling splitting increases with
temperature.\cite{ia,em} On the other hand, for a spin 1/2 system,
there is not excited state for each direction. The temperature
dependence on quantum tunneling depends on the spectrum of
tunneling splitting when coupling to the bath. Unfortunately, even
the bath effect on quantum tunneling in ground state is not
exactly known in the spin-boson model. Some discussion on
temperature effect given in Ref.\ \onlinecite{weiss} is for the
double-well system, but not for the spin-boson model.

\begin{acknowledgements}
This work was supported by a grant from the Natural Science
Foundation of China  under Grant No. 10575045.
\end{acknowledgements}

\end{document}